\documentclass[aps,pre,showpacs,twocolumn]{revtex4}

\usepackage{graphics}
\usepackage{epsfig}
\begin{document}
\title{Off-resonance field enhancement by spherical nanoshells}

\author{Andrey E. Miroshnichenko}
%\email{aem124@rsphysse.anu.edu.au}
\affiliation{Nonlinear Physics Centre, Research School of Physics and Engineering, Australian National University, Canberra ACT 0200, Australia}

\begin{abstract}
We study light scattering by spherical nanoshells consistent of metal/dielectric composites. We consider two geometries of metallic nanoshell with dielectric core, and dielectric coated metallic nanoparticle. We demonstrate that for both geometries the local field enhancement takes place out of resonance regions ("dark states"), which, nevertheless, can be understood in terms of the Fano resonance. At optimal conditions the light is stronger enhanced inside the dielectric material. By using nonlinear dielectric materials it will lead to a variety nonlinear phenomena applicable for photonics applications.  
\end{abstract}

\pacs{42.25.Bs, 42.25.Fx, 78.67.Bf}

\maketitle

\section{Introduction}

Electromagnetic properties of metallic nanoparticles have attracted a lot of attention during last decade due their ability to squeeze light at subwavelength level. Coherent excitation of conduction electrons on a metallic surface allows to enhanced optical field close to the surface of the particle. Such field enhancement was used for sensing, single molecule detection, surface enhanced Raman scattering, medicine, biology and other applications~\cite{maier}. The understanding of how metal geometries control the properties of surface plasmons will lead to a precise design of metallic nanoparticles for optimized optical field enhancement. One of the successful examples is metallic nanoshell structures with a dielectric core. For such structures plasmon hybridization theory was developed~\cite{epcrnjhpn:s:03, eppn:jcp:04}. In the framework of this theory plasmons of adjacent surfaces of complex metallic geometries may interact and, therefore, hybridize in analogy with the wave functions of a quantum system. The unique property of metallic nanoshell structures is the sensitivity of surface plasmons to the thickness of the metal shell.
It allows to design materials to match the required wavelength for a particular application in visible or near infrared regions.

But, existence of inherit optical losses of metals prevents them from successful application in photonics so far. Recently it was experimentally demonstrated that the conjugate geometry with metallic core and dye-doped coated dielectric may allow to overcome loss of surface plasmon resonances~\cite{nmagbambrsvmneesshestwu:n:09}. Moreover, it may lead to a first realization of a spaser (surface plasmon amplification by stimulated emission of radiation) nanoplasmonic counterpart of the laser, first predicted in Ref.~\cite{bdjsmi:prl:03}. Spaser-based nanolaser may become a key component for future nanophotonic technologies.

%%%%%%%%%%%%%%%%%%%%%%%%%%%%%%%%%%%%%
% FIGURE 1
%%%%%%%%%%%%%%%%%%%%%%%%%%%%%%%%%%%%%
\begin{figure}
\vspace{20pt}
\includegraphics[width=1.\columnwidth]{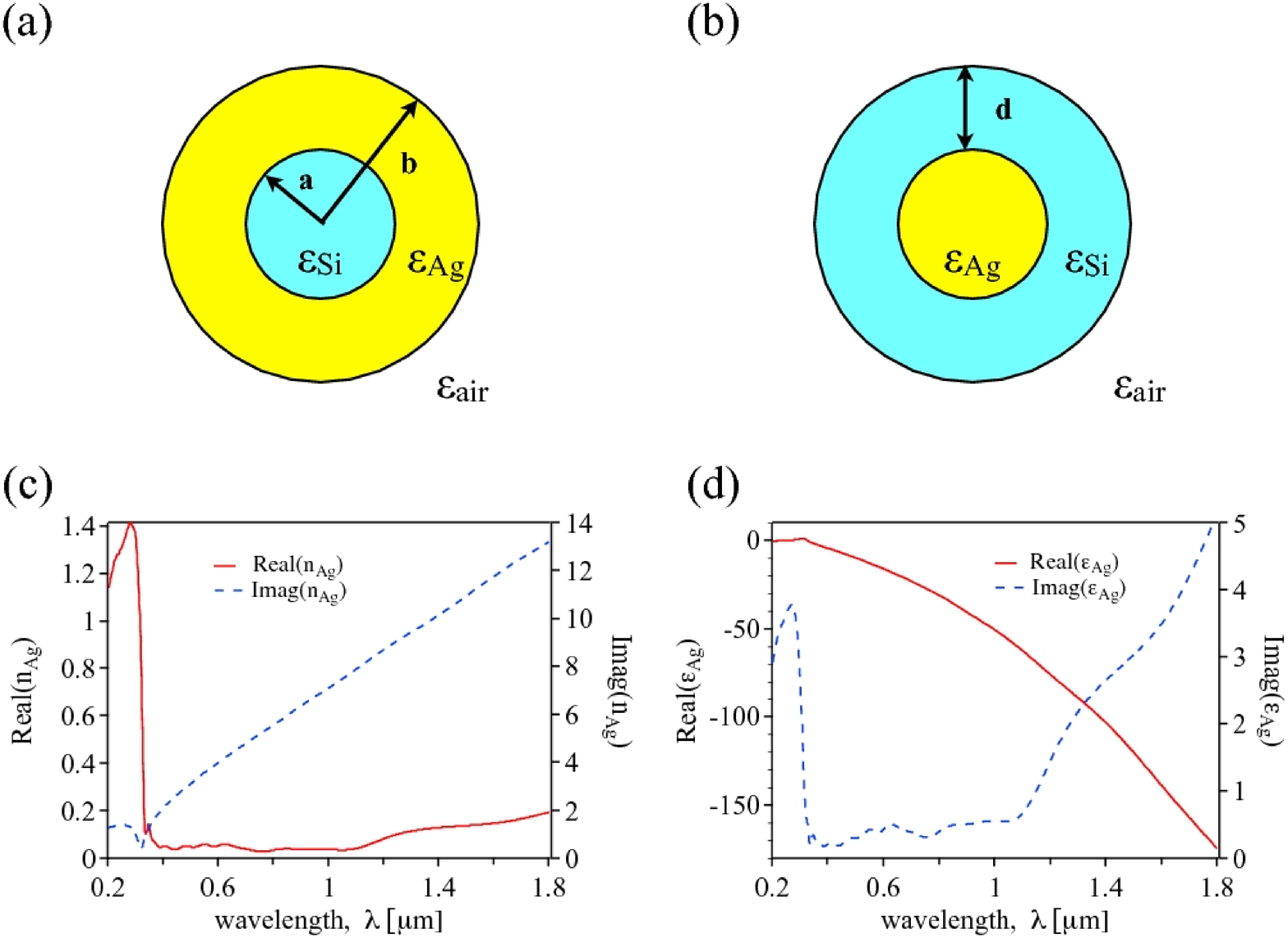}%
\caption{\label{fig:fig1} (Color online)
Schematic view of considered geometries: (a) silver nanoshell with silicon core, and (b) silicon coated silver nanoparticle. $a$ and $b$ are internal and external radii of the spherical shell, and $d=b-a$ is thickness of the coating layer. Plots (c) and (d) show extrapolated experimentally derived refractive index and permittivity of the silver~\cite{pbjrwc:prb:72}.
}
\end{figure}
%%%%%%%%%%%%%%%%%%%%%%%%%%%%%%%%%%%%%
% END OF FIGURE
%%%%%%%%%%%%%%%%%%%%%%%%%%%%%%%%%%%%%

In this paper we investigate both geometries of metallic nanoshell with dielectric core and dielectric coated metallic nanoparticle. For our analysis we consider silver/silicon composites. The main attention is paid to the local field enhancement inside and near the nanoparticle. One the common approaches to analyse field enhancement is by means of scattering resonances of a coated sphere. The Q-factor of a resonances is associated with amount of the field which can be stored inside the nanoparticle. But, our analysis reveals that maximal field enhancement does not necessary coincide with the enhanced scattering. On contrary, we demonstrate that in some case the maximal enhancement may correspond to resonant suppression of the scattering. These so-called "dark states" correspond to the strong local excitation which does not couple to the far field. Such behaviour can be interpreted in terms of the Fano resonances. The use of high refractive index dielectric allows to localize electric field inside the nonabsorbing layer. The enhanced light power inside silicon layer may potentially lead to a nonlinear response of the whole structure. This could be another approach to overcome losses of metals and use metallic nanoparticles in photonic applications.

\section{Light scattering by nanoshell particles}

Light scattering by a spherical particle is one of the most elaborated problems in electromagnetic theory. The analytical solution was found independently more than a hundred years ago by Lorenz~\cite{lvl:kdvss:90} and Mie~\cite{gm:ap:08}. Although this solution is know for a very long time and was analized from different perspectives it still contains some hidden results~\cite{aemsfavgbslyskmit:opn:08}. Aden and Kerker~\cite{alamk:jap:51} extended conventional Lorenz-Mie theory to the scattering by coated spheres with arbitrary parameters. The complete solution consists of infinite series of Bessel's function, which can be truncated depending on the particle size~\cite{bh}. But still it requires the use of computer to be fully analyzed. Up to now there were developed many special algorithms for fast and accurate calculations of Bessel's functions and their derivatives~\cite{links}.

\subsection{Fr\"ohlich modes}

Metallic nanoparticles are known to demonstrate localized surface plasmon resonances (LSPR), where the electro-magnetic field is greatly enhanced near the surface~\cite{bh}. 
According to Mie solution in the small particle limit $a\ll\lambda$, where $a$ is the particle radius and $\lambda$ is input wavelength, the condition for LSPR can be written as follows
\begin{eqnarray}\label{eq:Froelich1}
Re[\epsilon_m(\omega)]=-\frac{l+1}{l}\epsilon_o\;,
\end{eqnarray} 
where $\epsilon_m$ and $\epsilon_o$ are permittivities of the metal and surrounding dielectric, and $l=1,2,3\ldots$ is the mode number corresponding to dipole, quadrupole, octupole and so on resonances. One of the distinguish features of the surface modes is that the internal electric field does not have a radial node inside the metal~\cite{bh}. Due to dispersion of a metal the condition (\ref{eq:Froelich1}) can be satisfied for a number of frequencies for the same metal. Such frequencies are know in the literature as Fr\"ohlich frequencies~\cite{hfhp:ppss:55,bh,maier}, and corresponding modes are usually called as Fr\"ohlich modes. Although the absorption of metals can be relatively small $|Im[\epsilon_m(\omega)]|<1$, the absorption efficiency of the metallic nanoparticle at the Fr\"ohlich frequency can be effectively very large $Q_{\rm abs}\propto1/Im[\epsilon_m(\omega_F)]$. Thus, the scattering properties are greatly modified at LSPR. Recently, it was predicted that weakly absorbing particles may exhibit inverse hierarchy of optical resonances~\cite{mitbsl:prl:06}. It results in narrower resonances for higher order modes $l=1,2,3,\ldots$ with a possibility to store more energy inside due to larger Q-factor~\cite{mit:spj:84}.

We study light scattering by a spherical nanoshell consisting of metal-dielectric composite [see Fig.~\ref{fig:fig1}(a,b)]. We perform the comparative analysis of two different geometries. The first one is the metallic spherical shell with dielectric core in air [see Fig.~\ref{fig:fig1}(a)]. The second geometry is an opposite one with dielectric coated with metallic core [see Fig.~\ref{fig:fig1}(b)]. For our study without loss of generality we consider Si/Ag composites. For silicon we take refractive index to be $n=3.4$, and for silver we use experimental data~\cite{pbjrwc:prb:72} [see Fig.~\ref{fig:fig1}(c,d)] with small size Drude model correction~\cite{ukcvf:zphn:69,uk:jpmp:74}. 

The condition for excitation of the first-order surface mode $l=1$ for coated spherical nanoparticles can still be obtained analytically in the limit of small particle $b\ll\lambda$~\cite{bh} 
\begin{eqnarray}\label{eq:Froelich2}
(\epsilon_b+2)(\epsilon_a+2\epsilon_b)+\left(\frac{a}{b}\right)^3(2\epsilon_b-2)(\epsilon_a-\epsilon_b)=0\;,
\end{eqnarray}
where $\epsilon_a$ and $\epsilon_b$ are permittivities of the core and mantel of radii $a$ and $b$, respectively. According to this equation for the first geometry of silver nanoshell with silicon core $\epsilon_a = \epsilon_{\rm Si}$ and $\epsilon_b = \epsilon_{\rm Ag}(\omega)$ there are two Fr\"ohlich modes - surface-like with external medium, and void-like with internal dielectric. In general these two modes are excited at different frequencies, but there was some study demonstrating that under certain conditions they may overlap~\cite{tvtvvpfjga:prb:04,dwbpn:jcp:07}.  For the second geometry of the silicon coated silver nanoparticle in air with $\epsilon_a = \epsilon_{\rm Ag}(\omega)$ and $\epsilon_b = \epsilon_{\rm Si}$ there is single first-order surface mode. The Fr\"ohlich frequency of the coated silver lies in the region $-2\epsilon_{\rm Si}<\epsilon_{\rm Ag}(\omega_F)<-2$ depending on the thickness of the coating $d=b-a$. This result indicates that the coating allows to shift effectively the Fr\"ohlich frequency. 

In realistic systems surface modes can be detected via scattering resonances. According to Mie theory the scattering by spherical particles is described by scattering efficiency $Q_{\rm sca}$~\cite{bh}. In case of metallic nanoparticles the total extinction is the sum of scattering and absorption efficiencies
\begin{eqnarray}\label{eq:Q}
Q_{\rm ext} = Q_{\rm abs}+Q_{\rm sca}\;,
\end{eqnarray}
due to losses in the metal. As it was mentioned above, at Fr\"ohlich frequncies absorption by metallic nanoparticles can be greatly enhanced. For our studies we will use total extinction and absorption efficiencies, since it contains all needed information, and scattering efficiency can be derived from (\ref{eq:Q}). 

\subsection{Metallic nanoshell}

\begin{figure}
\vspace{20pt}
\includegraphics[width=1.\columnwidth]{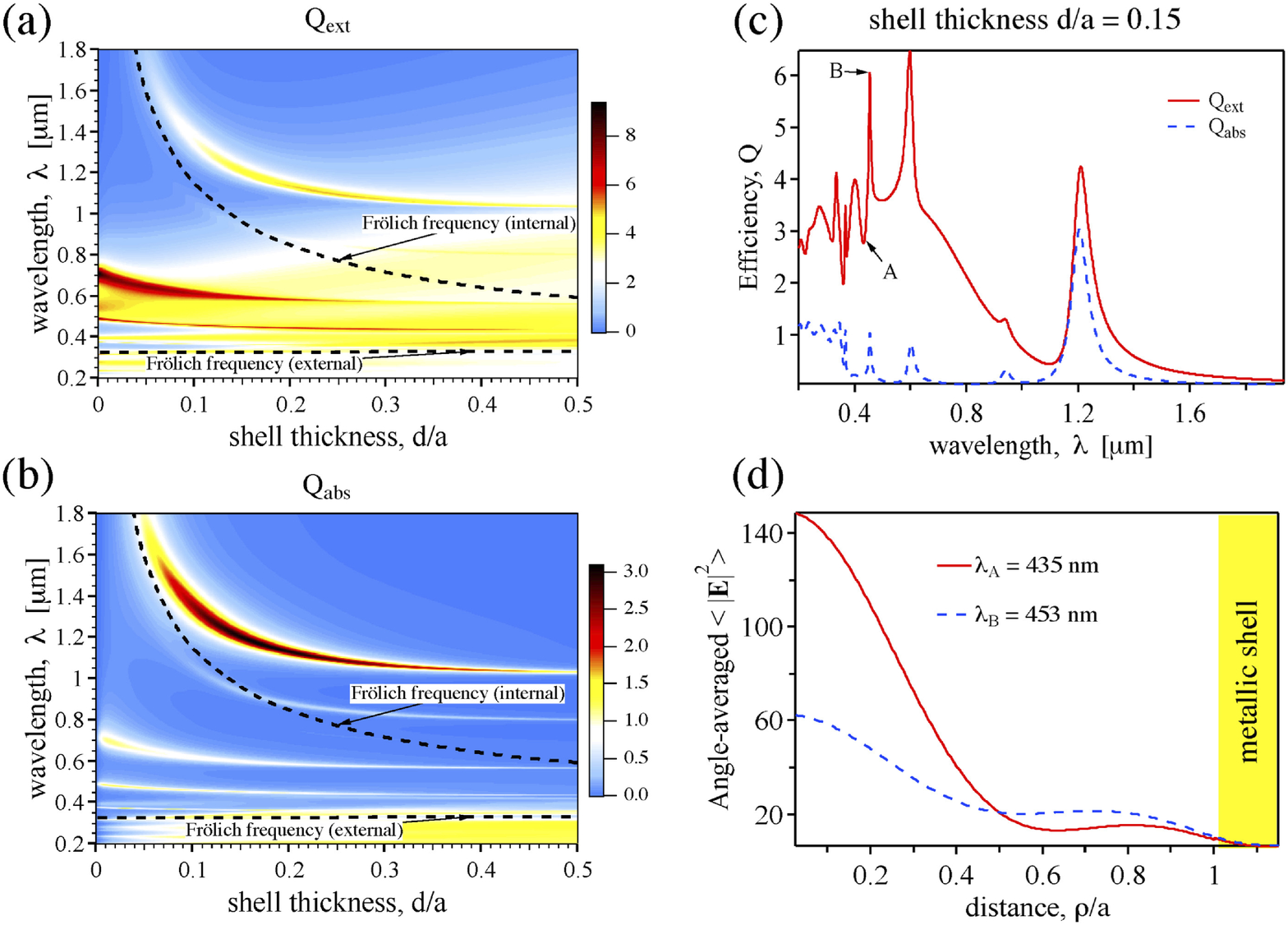}%
\caption{\label{fig:fig2} (Color online)
Scattering by silver nanoshell with silicon core of radius $a=100$ nm in air.
(a) extinction $Q_{\rm ext}$ and (c) absorption $Q_{\rm abs}$ efficiencies versus incident wavelength $\lambda$ and normalized shell thickness $d/a$. Dashed lines correspond to first-order Fr\"ohlich modes for small particle limit (\ref{eq:Froelich2}); (b) scattering for $d=15$ nm shell thickness; (d) angle-averaged electric field profile inside the nanoshell for two particular wavelength $\lambda_{\rm A,B}$ marked in plot (b). Grey aria indicates silver mantel.
}
\end{figure}

We start our consideration from the case of light scattering by silver nanoshell with silicon core. The results for this geometry are summarized in Fig.~\ref{fig:fig2}. The radius of the silicon core is fixed to $a=100$nm, and the silver shell thickness varies up to $d=50$nm. Both, extinction and absorption efficiencies saturate for larger shell thickness  because the optical skin depth of silver is less than $d_S<40$nm [see Fig.~\ref{fig:fig2}(a,b)]. The dashed lines in these plots indicate location of first-order void- and surface-like Fr\"ohlich modes according to eq.(\ref{eq:Froelich2}). They are in a good agreement for small shell thickness, where the perturbation analysis is still valid. Although they deviate from exact solution the general trend of the first-order mode dependence can be qualitatively grasped. In addition to the first-order modes there are higher order modes. The wavelength dependence of internal and external modes is different. Void- and surface-like modes exist below and above plasma frequency of silver ($\lambda_p\approx320$nm [see Fig.~\ref{fig:fig1}(c,d)]), respectively. For that particular geometry we do not observe overlapping of different modes, which will require the passing one of the modes through the plasma frequency~\cite{khugsc:prl:09}. Note here that maximal extinction and absorption resonances correspond to excitation of void-like internal modes~\cite{jjplasamhaaabap:jap:08,cwqlc:josa:08}, but take place for different modes number. Absorption is maximal for dipole internal mode, while extinction is maximal for quadrupole mode (so does the scattering efficiency) for the shell thickness less than the penetration depth $d<d_S$. This result indicates that the scattering by metallic nanoshell is stronger by accumulating light inside the dielectric core.

%%%%%%%%%%%%%%%%%%%%%%%%%%%%%%%%%%%%%
% FIGURE 2
%%%%%%%%%%%%%%%%%%%%%%%%%%%%%%%%%%%%%
%%%%%%%%%%%%%%%%%%%%%%%%%%%%%%%%%%%%%
% END OF FIGURE
%%%%%%%%%%%%%%%%%%%%%%%%%%%%%%%%%%%%%

There is a common believe that at stronger scattering resonances the light field is enhanced within or in the vicinity of the nanoparticle. But, in general, it's not true. To demonstrate that we study light scattering at fixed shell thickness $d=15$nm [see Fig.~\ref{fig:fig2}(c)]. By calculating the angle-averaged electric field intensity $\langle|\mathbf{E}|^2\rangle$ inside the nanoshell~\cite{tkslgs:ao:94} for this curve we found that the maximal field enhancement corresponds to the resonant suppression of the extinction [see Fig.~\ref{fig:fig2}(d)]. This kind of excitation can be called a "dark state" where the excited local field effectively does not emit any background radiation.  Meanwhile at the nearest octupole resonance the field enhancement is 3 times less. Such a behaviour can be easily understood in terms of the Fano resonance. Indeed, recently it was demonstrated that the light scattering by weakly absorbing particles exhibits asymmetric scattering lineshapes~\cite{mitsfaemavgysk:prl:08}, which is applicable to metal nanoshell structures as well. The asymmetry comes from the constructive and destructive interference of the incident and reemitted light of a Fr\"ohlich mode. Such asymmetric profiles were described for the first time by Ugo Fano in his seminal paper~\cite{uf:PR:61}, where he derived a formulae for such kind of resonances, which were named after him

%%%%%%%%%%%%%%%%%%%%%%%%%%%%%%%%%%%%%
% FIGURE 3
%%%%%%%%%%%%%%%%%%%%%%%%%%%%%%%%%%%%%
\begin{figure}
\vspace{20pt}
\includegraphics[width=1.\columnwidth]{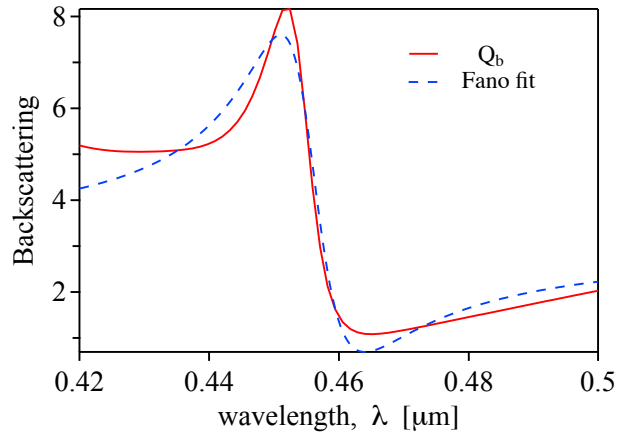}%
\caption{\label{fig:fig3} (Color online)
Calculated backscattering cross section $Q_b$ and its fitting with the Fano formulae (\ref{eq:Fano}) in the vicinity of wavelength $\lambda_{\rm A,B}$ in Fig.~\ref{fig:fig2}(c).
}
\end{figure}
%%%%%%%%%%%%%%%%%%%%%%%%%%%%%%%%%%%%%
% END OF FIGURE
%%%%%%%%%%%%%%%%%%%%%%%%%%%%%%%%%%%%%

\begin{eqnarray}\label{eq:Fano}
F(\epsilon)=\sigma_0\frac{(\epsilon+q)^2}{1+\epsilon^2}+\sigma_{\rm bg}\;,
\end{eqnarray}
with $\epsilon=2(E-E_R)/\Gamma$ where $E_R$ and $\Gamma$ are position and width of the resonance, $\sigma_0$ and $\sigma_{\rm bg}$ are normalized and background scattering, and $q$ is the asymmetry parameter. According to this formulae maximal and minimal scattering (constructive and destructive interference) take place at $E_{\rm max}=E_R+\Gamma/(2q)$ and $E_{\rm min}=E_R-\Gamma q/2$, respectively, while the actual resonance frequency is somewhere in between, depending on the sign of the asymmetry parameter $q$~\cite{review}. It allows us to immediately conclude that the maximal scattering might not necessarily correspond the excitation of the resonant mode, which explains not optimal field enhancement. To prove that we are dealing with the Fano resonance in this case we will fit the scattering with the Fano formulae (\ref{eq:Fano}). Since the extinction efficiency $Q_{\rm ext}$ is averaged among all possible directions and polarizations, the resonance properties are averaged as well. Therefore, we will look at the scattering at the particular direction, namely, at the backscattering cross section $Q_b$, which is polarization independent. The resulting backscattering in the vicinity of the resonance and its Fano fit is presented in Fig.~\ref{fig:fig3}. In the figure one can clearly see that the Fano formulae accurately describes positions of the local minima and maxima, indicating that this is indeed the Fano resonance. The discrepancy between actual and fitting curve comes from the fact that there are other resonances, which change asymptotic behaviour. Other resonances located very close one to another result in interaction between them and corresponding field redistribution. Based on these facts, one may conclude that resonant scattering and resonant field enhancement are not always related, and, in general, should be distinguish one from each other. Thus, for any particular application both properties could or should be optimized separately.

\subsection{Dielectric coated metallic nanoparticle}

\begin{figure}
\vspace{20pt}
\includegraphics[width=1.\columnwidth]{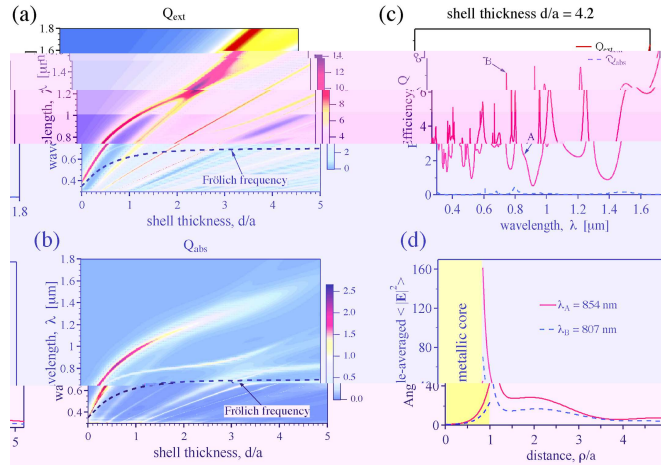}%
\caption{\label{fig:fig4} (Color online)
Scattering by silver nanoparticle of radius $a=50$ nm coated by silicon in air.
(a) extinction $Q_{\rm ext}$ and (c) absorption $Q_{\rm abs}$ efficiencies versus incident wavelength $\lambda$ and normalized shell thickness $d/a$. Dashed lines correspond to first-order Fr\"ohlich modes for small particle limit (\ref{eq:Froelich2}); (b) scattering for $d=210$ nm silicon coating; (d) angle-averaged electric field profile inside the nanoshell for two particular wavelength $\lambda_{\rm A,B}$ marked in plot (b). Grey aria indicates silver core.
}
\end{figure}

This behaviour is quite generic, and it's also applicable to the second geometry of silver nanoparticles coated with silicon. The results for this case are summarized in Fig.~\ref{fig:fig4}. We consider $a=50$ nm silver core with varying silicon mantel thickness. Extinction and absorption efficiencies in this case exhibit two types of resonances [see Fig.~\ref{fig:fig4}(a,b)], which can be treated in the frame of plasmon hybridization theory~\cite{epcrnjhpn:s:03}. One of them is associated with the LSPR of the silver core, and the second one corresponds Mie resonances of a silicon spherical particle. They can be clearly distinguish by their dependencies on the shell thickness $d$. According to the Mie theory of light scattering by dielectric spheres the positions of resonances can be characterized by the size parameter $x=2\pi r/\lambda$, where $r$ is the radius of the sphere. Thus, for larger spheres there is a linear redshift of Mie resonances with the radius of the sphere. Another types of resonares are LSPRs of the silver core, which lie below the plasma frequency, in contrast to the previous geometry [see Fig.~\ref{fig:fig2}(a,b)]. There is a nonmonotonous redshift of LSPRs with the shell thickness, predicted by eq.(\ref{eq:Froelich2}) [dashed line in Fig.~\ref{fig:fig4}(a,b)]. Note here, that the exact solution exhibits this much larger shift of resonances due to finite particle width, providing with wider possibilities of LSPRs tunability. Two types of resonances overlap at $d=130$ nm thickness of the coating layer leading to enhanced scattering by nanoshell particle [see Fig.~\ref{fig:fig4}(a,b)], although there is no enhancement of the light field inside the particle. Similar to previous case, we found that maximal filed enhancement takes place in out of resonance region, which is at least 2 times larger than at the near narrow resonance peak [see Fig.~\ref{fig:fig4}(c,d)]. Note here that the field is more strongly enhanced inside the silicon coating layer than inside the silver core. By utilizing nonlinear response of a dielectric such local field enhancement may lead for a variety of application in photonincs, including bistability and all-optical switching. 

%%%%%%%%%%%%%%%%%%%%%%%%%%%%%%%%%%%%%
% FIGURE 4
%%%%%%%%%%%%%%%%%%%%%%%%%%%%%%%%%%%%%
%%%%%%%%%%%%%%%%%%%%%%%%%%%%%%%%%%%%%
% END OF FIGURE
%%%%%%%%%%%%%%%%%%%%%%%%%%%%%%%%%%%%%

\section{Conclusions}

We studied light scattering by metal/dielectric spherical particles for two geometries. One of them is metallic nanoshell with dielectric core, and another is metallic nanoparticle coated by dielectric in air. In addition to the external LSPRs the first geometry supports internal or void-like resonances, where the electric field is more localized inside the dielectric core. We demonstrated that scattering resonances, in general, do not correspond to resonant field enhancement of the nanoparticle. We suggested that such a behaviour can be interpreted in therms of the Fano resonances, which are very common in the scattering by particles. It also implies that strong field enhancement for a particular applications should be sought independently from scattering resonances, although they are still linked to each other. 
In the case of dielectric coated metallic nanoparticle the field enhancement more than $160$ inside the dielectric mantel was achieved. The use of a nonlinear dielectric for coating in such geometry may open up new opportunities for achieving tunable sensors, or all-optical functionality for photonics applications.
 
\section*{Acknowledgments}
The author thanks Prof. Yuri Kivshar for useful discussions. 
The work has been supported by the Australian Research Council through the Discovery and Centre of Excellence projects.

\bibliography{mie_coated_minimal}

\end{document}